\documentclass[a4paper,11pt]{article}
\usepackage{pos}
\usepackage{cleveref}
\usepackage{lineno}
\usepackage{setspace}

\title{Development of an Autonomous Detection-Unit Self-Trigger for GRAND}

\author*[a]{Pablo Correa}
\author[a]{Jean-Marc Colley}
\author[b,c]{Tim Huege}
\author[d,e]{Kumiko Kotera}
\author[a]{Sandra Le Coz}
\author[a,d,f]{Olivier Martineau-Huynh}
\author[b]{Markus Roth}
\author[a,g]{Xishui Tian}

\onbehalf{for the GRAND collaboration}

\affiliation[a]{Sorbonne Université, Université Paris Diderot, Sorbonne Paris Cité, CNRS, Laboratoire de Physique Nucléaire et de Hautes Energies (LPNHE), 4 place Jussieu F-75252, Paris Cedex 5, France}
\affiliation[b]{Institute for Astroparticle Physics, Karlsruhe Institute of Technology, D-76021 Karlsruhe, Germany}
\affiliation[c]{Astrophysical Institute, Vrije Universiteit Brussel, Pleinlaan 2, 1050 Brussels, Belgium}
\affiliation[d]{Institut d’Astrophysique de Paris, CNRS UMR 7095, Sorbonne Universite, 98 bis bd Arago 75014, Paris, France}
\affiliation[e]{IIHE/ELEM, Vrije Universiteit Brussel, Pleinlaan 2, 1050 Brussels, Belgium}
\affiliation[f]{National Astronomical Observatories, Chinese Academy of Sciences, Beijing 100101, China}
\affiliation[g]{Department of Astronomy, School of Physics, Peking University, Beijing 100871, China}

\emailAdd{pablo.correa@lpnhe.in2p3.fr}

\abstract{One of the major challenges for the radio detection of extensive air showers, as encountered by the Giant Radio Array for Neutrino Detection (GRAND), is the requirement of an autonomous radio self-trigger. This work presents the current development of self-triggering techniques at the detection-unit level---the so-called first-level trigger (FLT)---in the context of the NUTRIG project. A second-level trigger (SLT) at the array level is described in a separate contribution. Two FLT methods are described, based on a template-fitting algorithm and a convolutional neural network (CNN). In this work, we compare the preliminary offline performance of both FLT methods in terms of signal selection efficiency and background rejection efficiency. We find that for both methods, ${\gtrsim}40\%$ of the background can be rejected if a signal selection efficiency of 90\% is required at the $5\sigma$ level.}

\FullConference{%
  10th International Workshop on Acoustic and Radio EeV Neutrino Activities (ARENA2024)\\
  11--14 June 2024\\
  The University of Chicago, Chicago, USA
}

\begin{document}
\maketitle

\section{Introduction}

The Giant Radio Array for Neutrino Detection (GRAND) \cite{Alvarez_2020,Kotera_2024} primarily aims at detecting ultra-high-energy (UHE; $E > 100$ PeV) tau neutrinos. Tau leptons produced in the charged-current interactions of Earth-skimming tau neutrinos with the Earth's crust can propagate into the atmosphere before decaying, thereby inducing nearly-horizontal (zeniths ${\sim}90^\circ$) extensive air showers. The transient (${\sim}$10 ns) radio emission produced by these air showers via geomagnetic and Askaryan effects is what GRAND aims to measure.


Currently, the main pathfinder array for GRAND is the GRANDProto300 (GP300) prototype, which is extensively described in \cite{Chiche_2024}. Upon completion, GP300 will consist of 230 detection units (DUs; each composed of a three-armed butterfly antenna and its front-end electronics) located in the Gobi desert near Dunhuang, Gansu province, China. However, in its final configuration, GRAND will comprise 20 subarrays of 10,000 DUs each, spread across the world for a maximal sky coverage. Since the footprint of very-inclined air showers covers several 10 $\rm km^2$ \cite{Aab_2018}, such GRAND10k arrays will have a sparse inter-DU spacing of 1 km, yielding a total instrumented area of 10,000 $\rm km^2$ per array. Consequently, to cope with such large scales, GRAND requires an autonomous radio trigger which is not only efficient, pure, and low cost, but also minimizes the data-communication bandwidth within the arrays.






The NUTRIG project is the first step towards the development of an autonomous radio trigger for large-scale radio arrays such as GRAND. NUTRIG consists of the following key aspects:
\begin{itemize}
    \item \textbf{First-level trigger (FLT) at the DU level.} To this day, radio self-triggers have mostly been using signal-over-threshold techniques to trigger on transient radio pulses \cite{Lautridou_2012,Asch_2008,Schmidt_2011,Charrier_2018}. This is also the case for the current DU trigger used in GP300, which we will name FLT-0 henceforth. With NUTRIG, we aim to complement the FLT-0 with a more elaborate FLT-1 algorithm, where we exploit the expected air-shower pulse shape to better reject transient radio frequency interference (RFI). While the nominal FLT-0 trigger rate is 1 kHz for GP300, we aim for an FLT-0$+$FLT-1 trigger rate of 100 Hz.
    
    \item \textbf{Second-level trigger (SLT) at the array level.} Instead of simply requiring a number of coincident FLTs within a certain time window, as currently done in GP300, with NUTRIG we are developing an SLT that utilizes the air-shower footprint features to better distinguish them from RFI coincidences. As such, we aim to reduce the nominal GP300 array-level trigger rate from 10 Hz to 1 Hz. A detailed description of the SLT is given in \cite{Kohler_2024}.
    
    \item \textbf{Modeling of air-shower radio emission between 50--200 MHz}. A precise description of the radio signal in this frequency range, which is where GRAND operates, will allow us to better exploit the air-shower features at the trigger level. See \cite{Gulzow_2024} for more details about the model.
\end{itemize}

In this work, we present the development of two candidate FLT-1 algorithms and their offline performance. First, we describe the construction of a dedicated database for the development and testing of these FLT-1 methods in Section \ref{sec:database}. The novel FLT-1 algorithms are subsequently described in Section \ref{sec:flt_algorithms}, and their offline performances are discussed in Section \ref{sec:results}. Finally, conclusions and future plans---for online trigger tests in particular---are outlined in Section \ref{sec:summary}.

\section{Database}
\label{sec:database}

\subsection{GRANDProto13}
\label{sec:gp13}

The NUTRIG-dedicated database partly relies on data recorded by the current 13-DU setup of GP300, named GRANDProto13 (GP13). Here, we give a very brief description of the GP13 setup; see \cite{Chiche_2024} for more details. Each solar-powered DU consists of a butterfly antenna with three orthogonal arms on top of a 3.5-m pole. These arms are oriented along the North-South axis (dipole), East-West axis (dipole), and upwards (monopole), which correspond to the X, Y, and Z channels of the DU, respectively. Signals captured by the antenna are first passed through a low-noise amplifier within the antenna nut, and sent via coaxial cables to a front-end electronics board (FEB) placed at the foot of the antenna. 

Within the FEB, the signals are processed through a variable-gain amplifier (fixed to 20 dB in this work) and a bandpass filter between 30--230 MHz before being digitized by a 14-bit analog-to-digital converter (ADC). The ADC sampling rate is 500 Msamples/s, and its operating range is $\pm 0.9$ V (13 bits for both positive and negative voltages). Digitized signals are then processed by a field-programmable gate array (FPGA), which contains the current FLT-0 trigger logic (Section \ref{sec:database_bkg}). After triggering, 4 central-processing-unit (CPU) cores are used for event building, and these events are then transferred to the central data-acquisition control room through a WiFi connection.

\subsection{Background}
\label{sec:database_bkg}

In order to create a database of RFI-background pulses, we use minimum-bias GP13 data recorded between 16 January -- 24 April 2024. Every 10 seconds, a trigger is forced in each DU, which records a trace of 1024 samples (2048 ns) for each of the three ADC channels; the combination of the three is what we define as a ``hit'' in this work. DUs 1032 and 1085 are excluded from this selection, as they exhibit abnormally high noise levels compared to the other 11 DUs during this period. The total number of recorded hits across all 11 DUs amounts to 6,695,504, corresponding to an integrated livetime of 13.71 s.

\begin{figure}[t]
    \centering
    \includegraphics[width=0.495\textwidth]{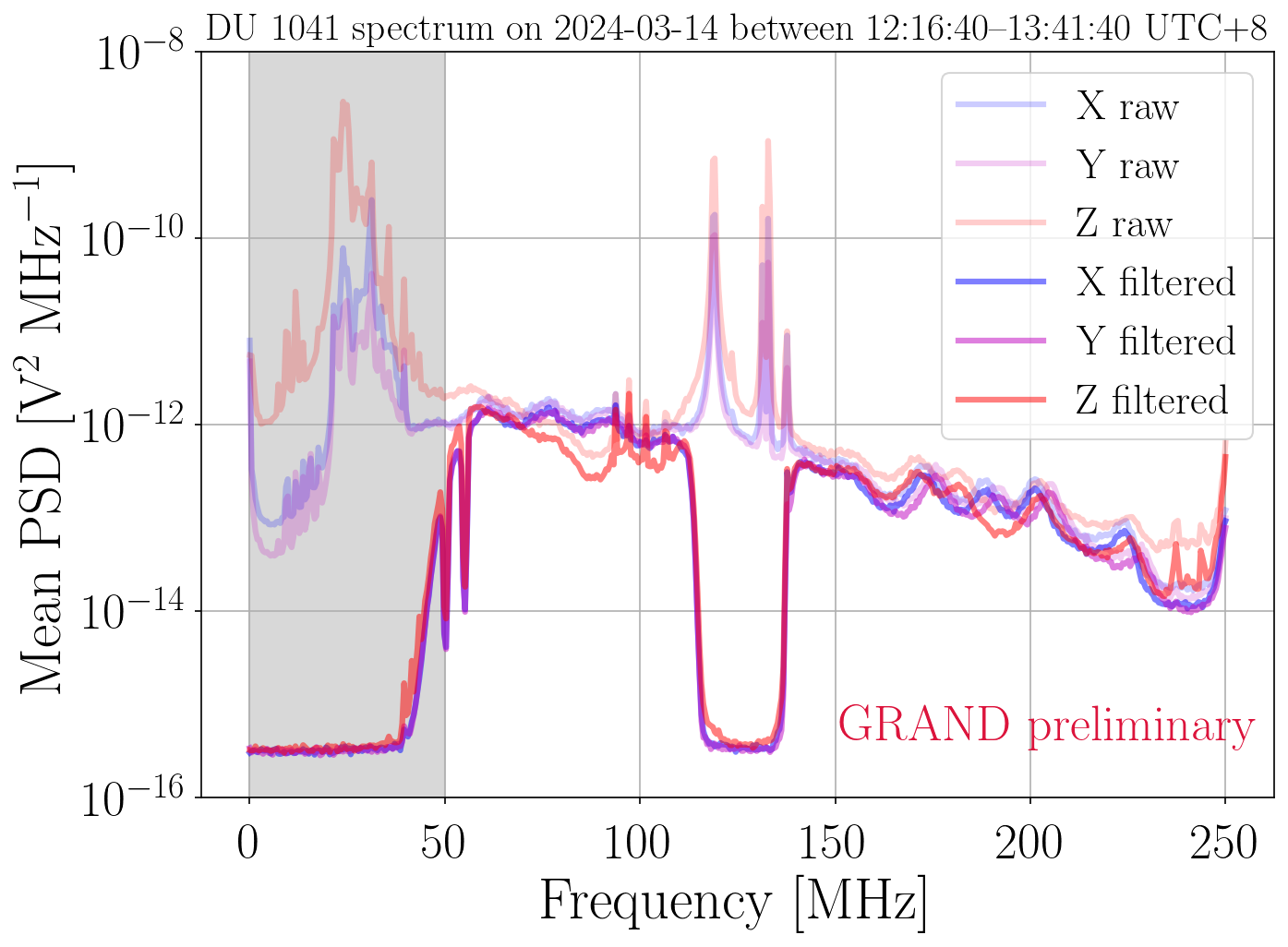}
    \includegraphics[width=0.495\textwidth]{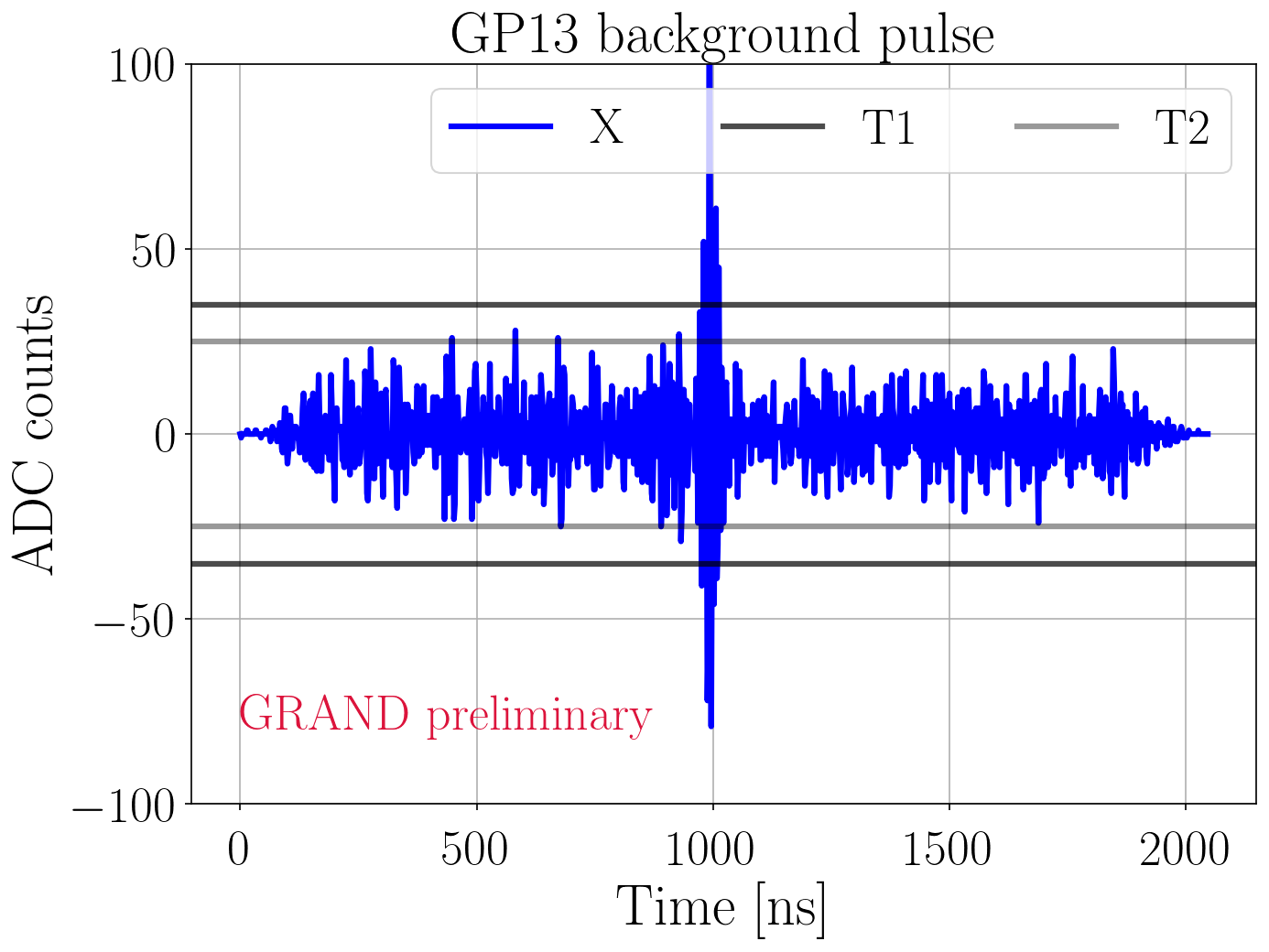}
    \caption{\textit{Left}: Mean power spectrum density (PSD) for DU 1041 of GP13 as obtained with minimum-bias data (see text for details). For all three channels, the transparent lines show the ``raw'' spectrum measured on site, while the opaque lines show the spectrum after applying the offline digital filters as described in the text. The shadowed area indicates the shortwave frequency range. \textit{Right}: An example of a GP13 background pulse (in the X channel) obtained in our database. The lines at $\rm T1 = 35$ ADC counts and $\rm T2 = 25$ ADC counts represent the thresholds of the offline FLT-0 algorithm used to identify transient pulses.}
    \label{fig:bkg_filter_flt0}
\end{figure}

As illustrated in the left panel of Fig.~\ref{fig:bkg_filter_flt0}, the GP13 spectrum is dominated by shortwaves below 50 MHz and aeronautic communication lines between 113.5--139.5 MHz. In order to mitigate these backgrounds, Butterworth filters are applied to the minimum-bias data in these bands. In addition, two notch filters are applied to kill prominent emission lines that occasionally appear at 50.2 MHz and 55.5 MHz. Note that Gaussian windowing is applied to the data before filtering, in order to avoid boundary effects. The filtered spectra are also shown in the left panel of Fig.~\ref{fig:bkg_filter_flt0}.

After filtering the data, we approximate the FLT-0 double-threshold trigger logic that is currently integrated in the GP13 firmware. For a trace $V$ to be flagged as a pulse, $|V|$ needs to yield ${\geq}2$ samples above the ``signal threshold'' T1, and ${\geq}6$ samples above the ``background threshold'' $\rm T2 < T1$. A quiet time of 50 ns is enforced before the first T1 crossing, and two consecutive samples above T1 and T2 cannot be more than 50 ns apart. To ensure that the pulse is completely contained within the trace, we exclude the first and last 100 samples of the trace in this treatment. The time of the first T1 crossing is defined here as the FLT-0 trigger time, $t_0$. If there are multiple pulses in a trace, we only tag the first pulse as an FLT-0 trigger.

The FLT-0 is applied exclusively to channels X and Y; channel Z is ignored in this work since it has a distinct RF chain leading to a larger dispersion of transient pulses. In order to test the performance of the FLT-1 methods down to low signal-to-noise ratios (SNRs), we choose $\rm T1 = 35$ ADC counts and $\rm T2 = 25$ ADC counts, lower than actual values used online. 14.8\% of the hits pass the aforementioned criteria and are identified as ``background pulses''; an example is shown in the right panel of Fig.~\ref{fig:bkg_filter_flt0}. Hits that do not pass the FLT-0 criteria are tagged as ``stationary background''. Finally, background pulses are selected randomly to create a background training sample and a background testing sample containing 10,000 hits each.

\subsection{Signal}
\label{sec:database_sig}

For the database of air-shower pulses, we start from electric-field simulations performed with ZHAireS \cite{Alvarez-Muniz_2011} for the complete GP300 array\footnote{Note that these simulations are not performed using the final GP300 layout, which was only approved recently \cite{Chiche_2024}.}. A total of 12,500 proton-induced air showers were simulated, covering primary energies uniformly in $\log_{10}(E/{\rm eV}) \in [16.5,18.6]$, zeniths $\theta \in [30.6^\circ,87.3^\circ]$ following a $\log_{10} (1/\cos \theta)$ distribution, and azimuths uniformly in $\phi \in [0^\circ,360^\circ]$. Using the dedicated software package GRANDlib \cite{Alves_2024}, the simulated electric fields are then convoluted with the antenna response and subsequently processed through the entire radio-frequency (RF) chain up to the ADC (Section \ref{sec:gp13}).

\begin{figure}[t]
    \centering
    \includegraphics[width=0.508\textwidth]{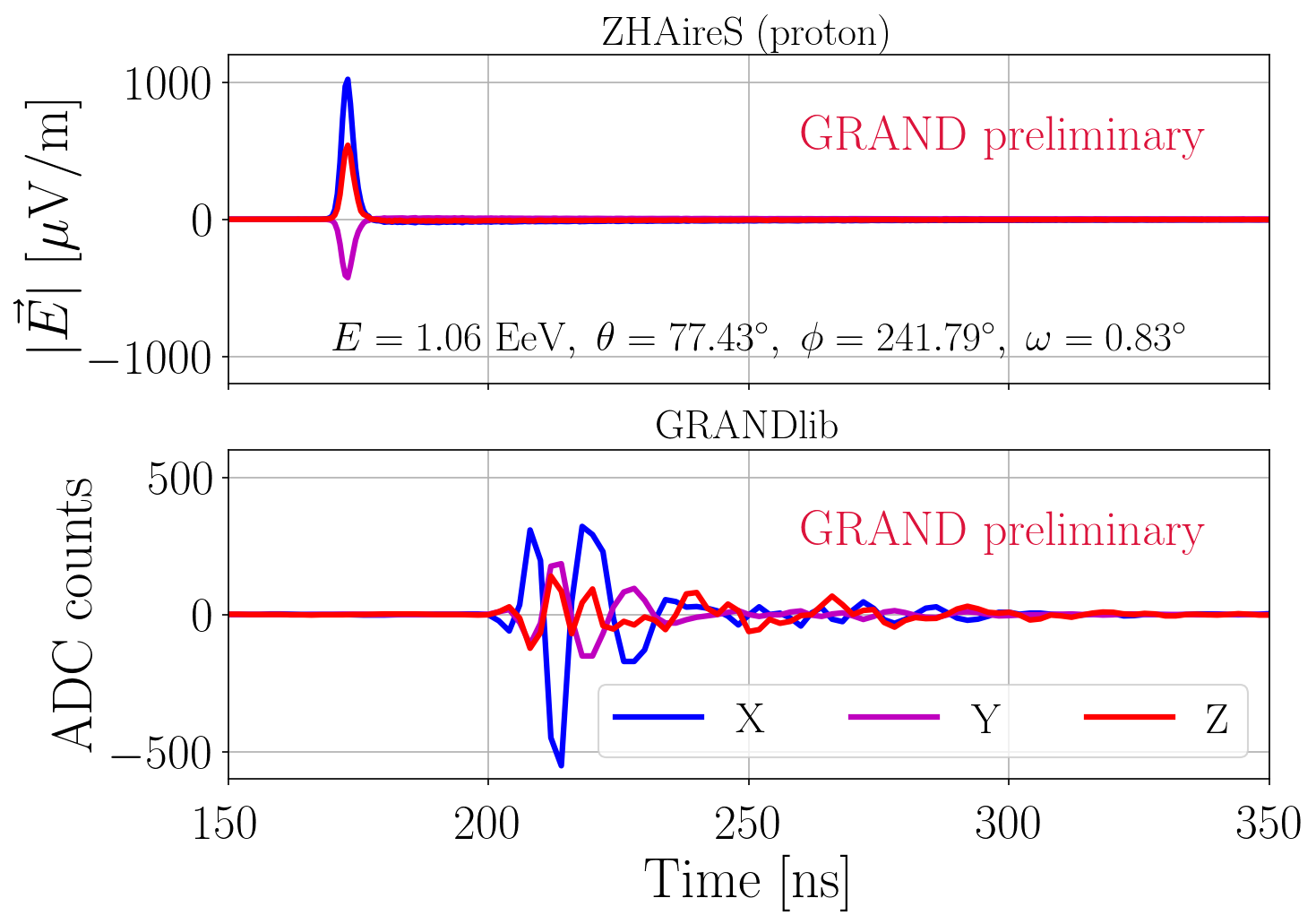}
    \includegraphics[width=0.482\textwidth]{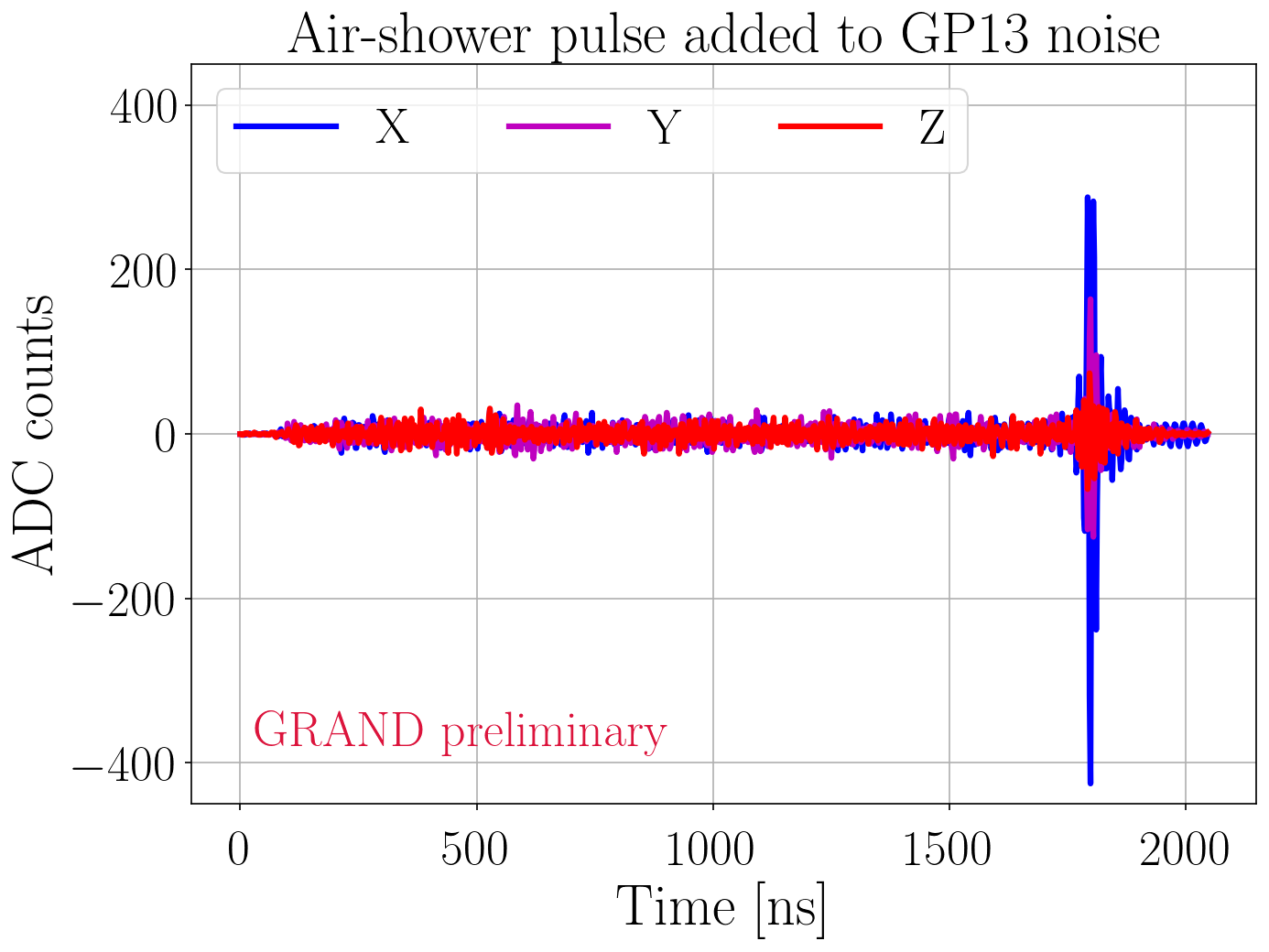}
    \caption{\textit{Top left}: Electric-field pulse simulated with ZHAireS, for the three channels of a DU located near the Cherenkov cone of a proton-induced air shower; shower properties are also given. \textit{Bottom left}: The same pulse at the ADC level after being processed through the GP300 RF chain with GRANDlib. \textit{Right}: A similar air-shower pulse at ADC level, but added to a minimum-bias hit of GP13 without background pulse.}
    \label{fig:sig_pulse}
\end{figure}

At the ADC level, the simulated air-shower pulses are added to minimum-bias data of GP13 that does not contain background pulses. As such, for each simulated DU, we select a random hit from the stationary background obtained in Section ~\ref{sec:database_bkg}. Next, for each channel, we add the pulse in a random position of the noise trace---excluding the first and last 100 samples of the trace---to mimic the fact that the background pulses in minimum-bias data can occur anywhere in the trace. After that, we perform the same treatment as done for the background, i.e., we filter the traces and apply the FLT-0 algorithm\footnote{Note that for the signal database used in the offline SLT study of \cite{Kohler_2024}, we apply more stringent FLT-0 thresholds---namely $\rm T1 = 55$ ADC counts and $\rm T2 = 35$ ADC counts---that roughly correspond to an average 1 kHz trigger rate.} to them (only for channels X and Y). Finally, from the entire pool of ``signal pulses'' that pass the FLT-0, we randomly select 10,000 simulated hits for a signal training sample, and another 10,000 for a signal testing sample. Note that here, we weight the simulated events with $E^{-3}$ to not bias ourselves to high-SNR signal pulses in the selection.

\section{FLT-1 Algorithms}
\label{sec:flt_algorithms}

\subsection{Template Fitting}
\label{sec:flt_template}

As a first FLT-1 method, we consider a classical template-fitting approach, inspired by previous offline template analyses \cite{Anker_2020,Henrichs_2023,Mitra_2023}. Here, we aim to exploit the shape of the air-shower pulses at ADC level to distinguish them from background pulses. Since the template-fitting FLT-1 algorithm relies on the FLT-0 trigger time $t_0$, as outlined below, it is only applied to the X and Y polarizations of a hit. 


For the template selection, we start from the same air-shower simulations of Section \ref{sec:database_sig} at ADC level (without added noise and ignoring the Z channel). Since the primary energy only affects the signal amplitude and not its shape, we only consider simulations for primary energies $E > 1$ EeV, and traces where the maximum of the trace exceeds 75 ADC counts. We then bin the simulated air-shower pulses in a $10 \times 10$ histogram covering $\theta \in [30.6^\circ,87.3^\circ]$ and $|\omega - \omega_c|/\omega_c \in [0,2]$, with $\omega$ the opening angle w.r.t.~the shower axis, and $\omega_c$ the Cherenkov angle. From each non-empty bin we randomly select one template to maximally cover possible differences in air-shower pulse shapes, resulting in a library of 96 templates. 

The FLT-1 template-fitting algorithm consists of the following four steps:
\begin{enumerate}
    \item \textbf{Fit the pulse-peak time.} For each polarization of interest, we first compute the cross correlation of a template $T$ with an input trace $V$,
    \begin{equation}
        \rho_T(\tau) = \left| \int \frac{T(t)}{\mathrm{RMS}[T(t)]}\, \frac{V(t+\tau)}{\mathrm{RMS}[V(t+\tau)]}\, {\rm d}t \right|,
    \end{equation}
    which is normalized between $[0,1]$. The correlation is performed in a window $\tau \in [t_0-{\rm 30~ns},t_0+{\rm 30~ns}]$ around the FLT-0 trigger time $t_0$, to account for possible jitter and pulse peaks after the first T1 crossing (Section \ref{sec:database_bkg}). The best pulse-peak time is then defined as the time where the correlation is maximized, $\hat{t}_{p,T} = {\rm argmax_\tau}\, |\rho_T(\tau)|$.

    \item \textbf{Fit the template amplitude.} In this step, we perform a $\chi^2$ minimization of the form
    \begin{equation}
        \chi^2_T(\kappa) = \sum_i | V(t_i) - \kappa T(t_i) |^2,
    \end{equation}
    where $t_i \in [\hat{t}_{p,T}-{\rm 20~ns},\hat{t}_{p,T}+{\rm 60~ns}]$. This fit window considers the most prominent part of the template around the pulse peak, allowing us to disfavor background-RFI transients which can be more extended than air-shower pulses. 

    \item \textbf{Find the best cross correlation.} We repeat steps 1 \& 2 for all templates in our sample. The optimal template is then given by the one with the smallest $\chi^2$, $\hat{T} = {\rm argmin}_T\, \chi^2_T$. The corresponding best cross correlation is given by $\hat{\rho} = \rho_{\hat{T}}(\hat{t}_{p,\hat{T}})$.

    \item \textbf{Compute a test statistic.} For each considered polarization $P$, step 3 yields a best cross correlation $\hat{\rho}_P$. We define our test statistic as the maximum of all cross correlations, $\mathrm{TS} = \max_P \hat{\rho}_P$. In this work, $\mathrm{TS} = \max \lbrace {\rm \hat{\rho}_X, \hat{\rho}_Y} \rbrace$.
    
\end{enumerate}




\subsection{Convolutional Neural Network}
\label{sec:flt_cnn}

The second FLT-1 algorithm considered in this study is based on a convolutional neural network (CNN) constructed within the Keras/Tensorflow framework. The details of the CNN method have been described extensively in \cite{LeCoz_2023}, although we note that in this work we are using two convolution layers instead of three. The CNN is trained using the dedicated background and signal training samples obtained in Sections \ref{sec:database_bkg} and \ref{sec:database_sig}. The output of the CNN is a score in $[0,1]$, which can be used to classify hits as background (smaller score) and signal (larger score). If we impose that background (resp.~signal) should be classified with a score below (resp.~above) 0.5, the CNN reaches an accuracy of ${\sim} 75\%$. 

\section{Offline FLT-1 Results}
\label{sec:results}

\begin{figure}[t]
    \centering
    \includegraphics[width=0.495\textwidth]{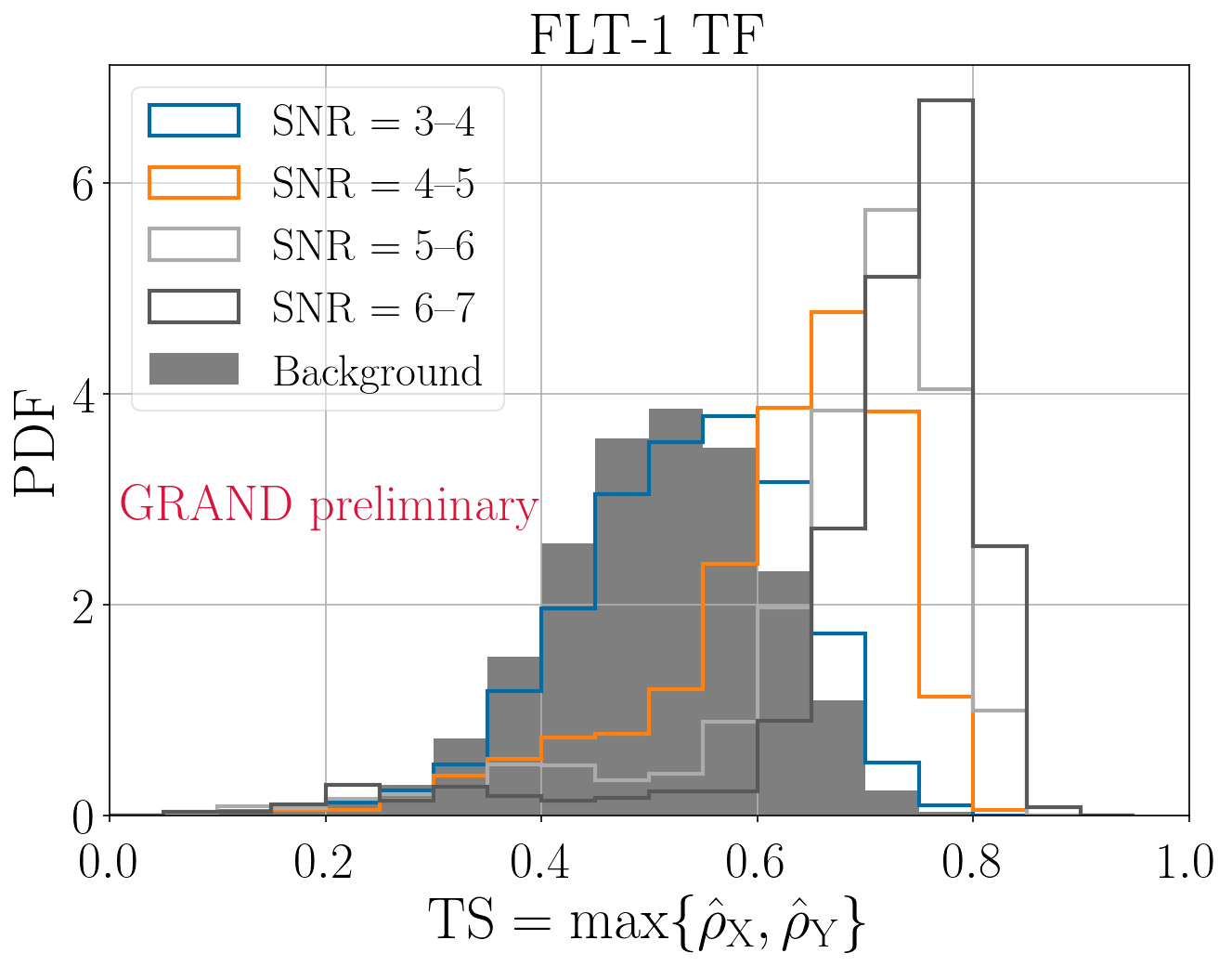}
    \includegraphics[width=0.495\textwidth]{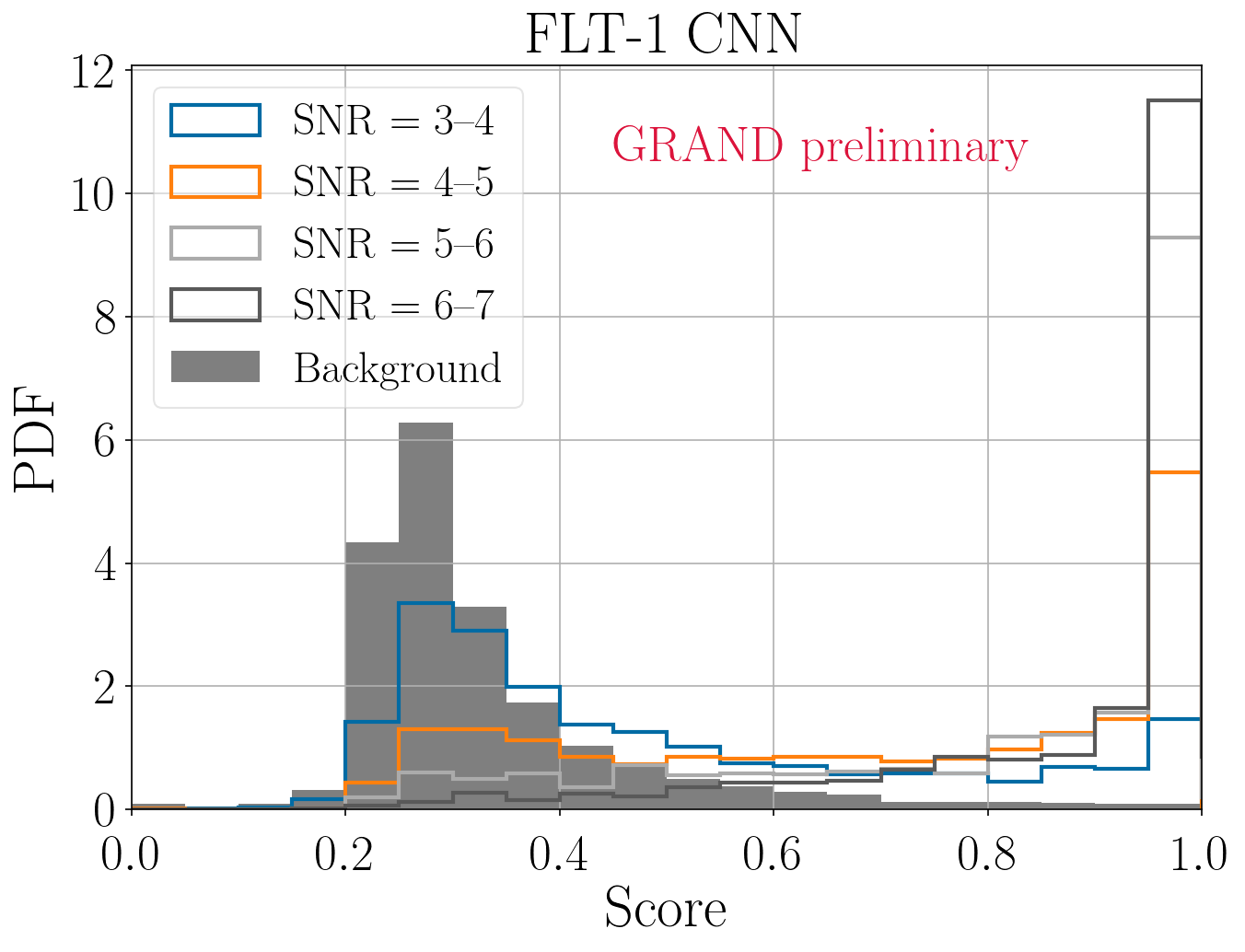}
    \caption{\textit{Left}: Distribution of the test statistic TS obtained with the FLT-1 template-fitting (TF) algorithm, for background (filled histogram) and different SNR ranges of the signal (empty histograms). \textit{Right}: Similar as the left panel, but now showing the score distributions obtained with the FLT-1 CNN algorithm.}
    \label{fig:flt_ts_distributions}
\end{figure}

Figure~\ref{fig:flt_ts_distributions} shows the distributions of the test statistic and score obtained with the FLT-1 template-fitting and CNN methods, respectively. The distributions are obtained using the background and signal testing samples of Sections \ref{sec:database_bkg} and \ref{sec:database_sig}. We divide the signal into different SNR ranges, where we define the SNR as the maximum of the simulated air-shower pulse divided by the RMS of the noise trace it is embedded in. We find that for both methods, the separation of signal and background increases with SNR, although the CNN FLT-1 generally has a more powerful signal-background separation than the template-fitting FLT-1.

\begin{figure}[t]
    \centering
    \includegraphics[width=0.495\textwidth]{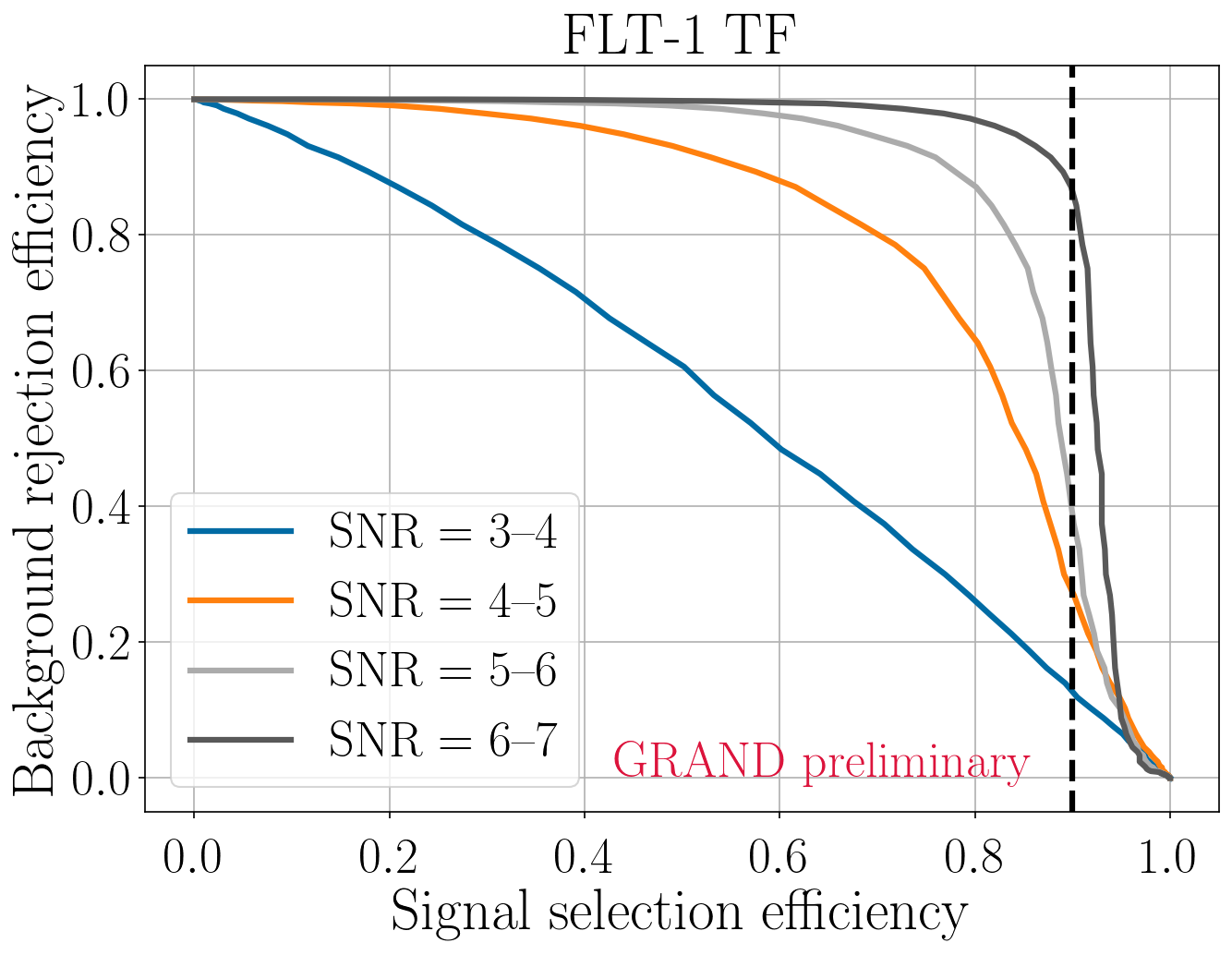}
    \includegraphics[width=0.495\textwidth]{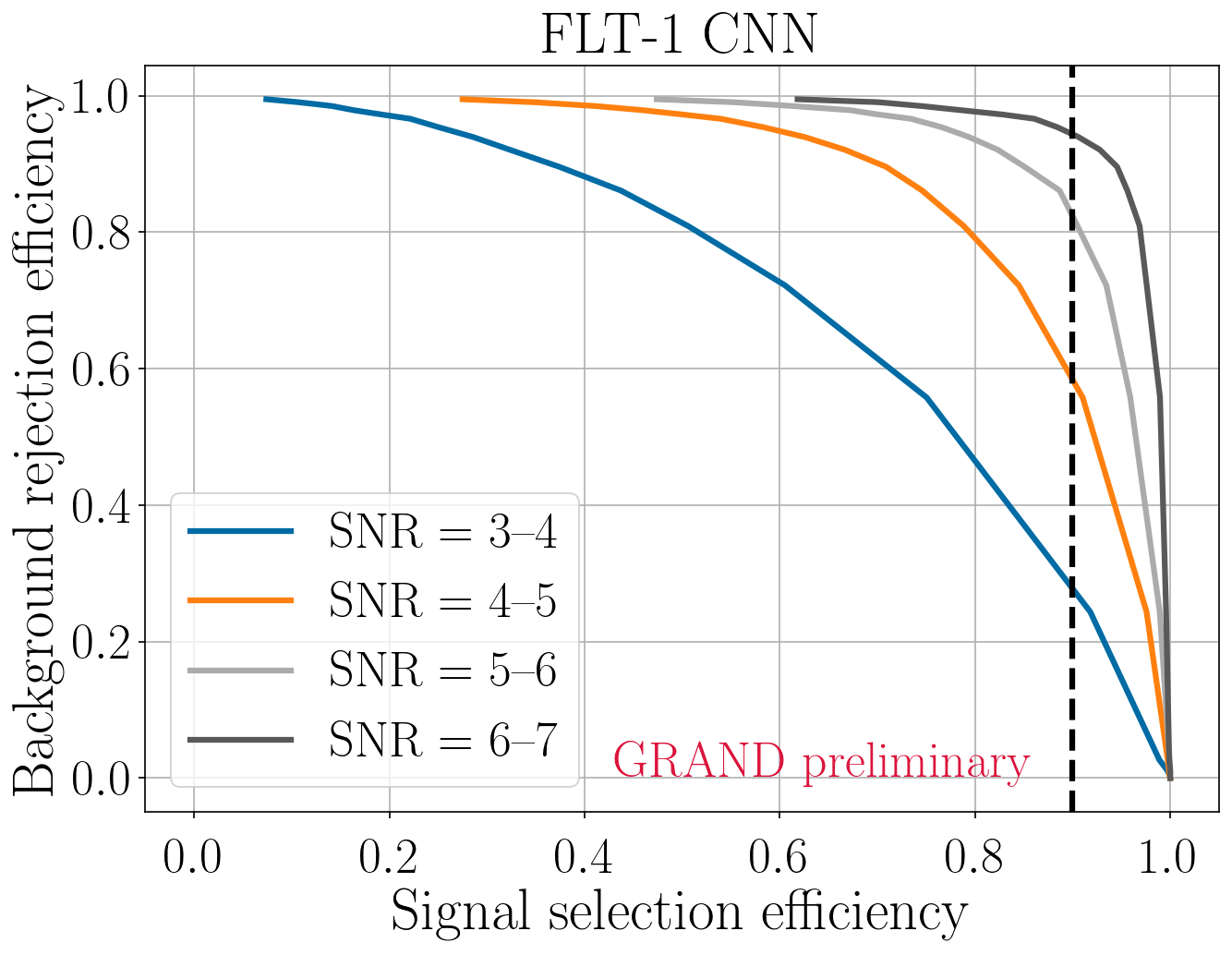}
    \caption{Background rejection efficiency as a function of signal selection efficiency, for both the FLT-1 template-fitting (TF; \textit{left}) and CNN (\textit{right}) algorithms, obtained using the distributions of Fig.~\ref{fig:flt_ts_distributions}. In each panel, the solid lines correspond to the efficiencies for different SNR ranges of the signal, and the dashed line corresponds to a signal selection efficiency of 90\%.}
    \label{fig:flt_offline_performance}
\end{figure}

Similar conclusions can be drown from Fig.~\ref{fig:flt_offline_performance}, which illustrates the offline performance of the two FLT-1 algorithms in terms of background rejection efficiency and signal selection efficiency. Overall, the CNN FLT-1 yields better background rejection efficiencies than the template-fitting FLT-1 for the same signal selection efficiency, for all considered SNR ranges. Nevertheless, for a signal selection efficiency of 90\%, both methods yield a background rejection efficiency $\gtrsim$40\% for signals with an $\rm SNR > 5$.

\section{Summary and Outlook}
\label{sec:summary}

In this work, we introduced two novel FLT-1 self-trigger methods at the DU level in the context of the NUTRIG project. While one FLT-1 method uses a classic template-fitting approach, the other is based on machine-learning techniques in the form of a CNN. We constructed a dedicated database to test these algorithms, and tested their offline performance. We found that generally, the CNN FLT-1 obtains a better signal selection efficiency and purity compared to the template-fitting FLT-1. Nevertheless, for a signal selection efficiency of 90\% at the $5\sigma$ level, both methods yield a background rejection efficiency of at least $\sim$40\%.

The next step is to port these FLT-1 algorithms to the CPU on the GP300 FEB. As such, we will be able to test the FLT-1 algorithms in controlled online conditions at the LPNHE test bench in Paris \cite{Correa_2023}. More specifically, we aim to compare the online performance of the FLT-0$+$FLT-1 combination with the nominal FLT-0 used in GP300. After that, we will test the FLT-0$+$FLT-1 setup in more realistic conditions at the dedicated GRAND@Nançay prototype \cite{Correa_2023}. Finally, in the mid-term future we plan to test the complete FLT+SLT methodology of NUTRIG at GP300, which will consist of 80 DUs by the end of 2024 \cite{Chiche_2024}.

\section*{Acknowledgments}
This work is part of the NUTRIG project, supported by the Agence Nationale de la Recherche (ANR-21-CE31-0025; France) and the Deutsche Forschungsgemeinschaft (DFG; Projektnummer 490843803; Germany). In addition, this work is supported by the CNRS Programme Blanc MITI (2023.1 268448 GRAND; France) and the Programme National des Hautes Energies (PNHE; France)
of CNRS/INSU with INP and IN2P3, co-funded by CEA and CNES. Computations were performed using the resources of the CCIN2P3 Computing Center (Lyon/Villeurbanne, France), a partnership between CNRS/IN2P3 and CEA/DSM/Irfu.

\small
{\setstretch{1}
\bibliographystyle{ICRC}
\bibliography{references}}

\end{document}